\documentclass[a4paper, amsfonts, amssymb, amsmath, reprint, showkeys, showpacs, nofootinbib, twoside, superscriptaddress,aps,prd]{revtex4-1}
\usepackage[english]{babel}
\usepackage[utf8]{inputenc}
\usepackage{comment}
\usepackage[colorinlistoftodos, color=green!40, prependcaption]{todonotes}
\usepackage{appendix}
\usepackage{amsthm}
\usepackage{mathtools}
\usepackage{physics}
\usepackage{xcolor}
\usepackage{graphicx}
\usepackage{adjustbox}
\usepackage{placeins}
\usepackage[T1]{fontenc}
\usepackage{lipsum}
\usepackage{csquotes}

\usepackage{ulem}
\usepackage{hyperref}

\begin{document}
\title{Strong deflection of massive particles in spherically symmetric spacetimes}

\author{Fabiano Feleppa}
    \email[Correspondence email address: ]{ffeleppa@unisa.it}
     \affiliation{Dipartimento di Fisica “E.R. Caianiello”, Università di Salerno, Via Giovanni Paolo II 132, I-84084 Fisciano, Italy}
    \affiliation{Istituto Nazionale di Fisica Nucleare, Sezione di Napoli, Via Cintia, 80126, Napoli, Italy}

    \author{Valerio Bozza}
    \email[Correspondence email address: ]{vbozza@unisa.it}
    \affiliation{Dipartimento di Fisica “E.R. Caianiello”, Università di Salerno, Via Giovanni Paolo II 132, I-84084 Fisciano, Italy}
    \affiliation{Istituto Nazionale di Fisica Nucleare, Sezione di Napoli, Via Cintia, 80126, Napoli, Italy}

    \author{Oleg Yu.\ Tsupko}
    \email[Correspondence email address: ]{tsupkooleg@gmail.com}
    \affiliation{ZARM, University of Bremen, 28359 Bremen, Germany} \affiliation{\mbox{Institut für Theoretische Physik, Goethe Universität, Max-von-Laue-Strasse 1, 60438 Frankfurt, Germany}}

\begin{abstract}
Near a gravitating compact object, massive particles traveling along timelike geodesics are gravitationally deflected similarly to light. In this paper, we study the deflection angles of these particles in the strong deflection limit. This analytical approximation applies when particles in unbound orbits approach the compact object very closely, circle around it at a radius close to that of the unstable circular orbit, and eventually escape. While previous studies have provided results for particular metrics, we offer a general solution applicable to any static, spherically symmetric and asymptotically flat spacetime. After briefly reviewing the exact expression for the deflection angle of massive particles, we present a strong deflection limit analysis for this general case. The developed formulas are then applied to three particular metrics: Schwarzschild, Reissner-Nordström and Janis-Newman-Winicour.
\end{abstract}


\maketitle

\section{Introduction}

The deflection of particles by gravity is primarily studied in the case of light rays. Measuring the deflection of starlight during a total solar eclipse provided the first experimental proof of general relativity. Since then, gravitational lensing has become a crucial tool in astrophysics, allowing one to investigate a wide range of phenomena. In most observational scenarios, the bending of light by gravity is very small. Therefore, the well-known Einstein formula for the deflection angle, which assumes a weak deflection, is usually sufficient for describing these lensing events.

On the other hand, the groundbreaking imaging of the supermassive black hole in M87 \cite{L1,L2,L3,L4,L5,L6} and the black hole at the center of our Galaxy \cite{L12,L13,L14,L15,L16,L17} have marked the beginning of a new chapter in the study of gravitational lensing, opening up possibilities to test general relativity beyond the weak deflection limit. When light rays are deflected by extremely compact gravitating objects, the bending can indeed become significant. Theoretical studies on the deflection of light by compact objects have also a long history. In 1959, Darwin \cite{Darwin1959} studied the motion of particles in a Schwarzschild spacetime. Notably, he developed a logarithmic approximation, now known as the strong deflection limit, to describe the deflection of light rays moving near the photon sphere --- the boundary encompassing all unstable circular orbits of photons with any possible inclinations. The strong deflection limit analysis proposed by Darwin was much later generalized to any spherically symmetric solution to Einstein field equations in Ref.\ \cite{Bozza2002}, as well as to axisymmetric solutions \cite{Bozza2003}. Over the years, a multitude of both analytical and numerical studies extending beyond the weak deflection limit approximation have been published (see, e.g., Refs.\ \cite{Hilbert-1917, Atkinson1965, Misner1973, Luminet1979, Ohanian1987, Ellis2000, Frittelli2000, Claudel2001, Bozza2001, Hasse2002, Beloborodov-2002, Perlick2004-review, Perlick2004, Amore-2006, Amore-2007, Iyer2007, Hackmann-2008a, Hackmann-2008b, Keeton2008, Tsupko2008, Majumdar2009, Bozza2010, Tarasenko2010, Eiroa2011, Wei2012, Zhang2015, Semerak-2015, Alhamzawi2016, Tsukamoto2016, Aldi-Bozza-2017, Dai2018, Aratore2021, Kuang2022, Cieslik-2022, Cieslik-2023, Aratore-Bozza-2024, Claros-Gallo-2024}). Additionally, several recent studies have focused on higher-order images specifically in the form of photon rings (see, e.g., Refs.\ \cite{Gralla2019, Johnson-2020, Gralla2020, Lupsasca2020, Gralla-Lupsasca-2020, Wielgus-2021, Broderick-2022, Ayzenberg-2022, Guerrero-2022, BK-Tsupko-2022, Tsupko-2022, Eichhorn-2023, Broderick-Salehi-2023, Kocherlakota-2024-1, Kocherlakota-2024-2, Aratore-Tsupko-Perlick-2024}). 

Analogous to photons, massive particles are also deflected in the presence of a gravitational field. The motion of massive test particles in black hole-like spacetimes has a long history as well \cite{Hagihara1931, Darwin1959, Misner1973, Darwin1961, Bogorodsky1962, Mielnik1962, Metzner1963, Novikov1971, Weinberg1972, Silverman-1980, Chandrasekhar1983, Rodríguez1987, Rodríguez1987II, Zakharov-1988, Zakharov-1994, Landau1993}. Over the past two decades, numerous works have also focused on calculating the deflection angle of massive particles by compact objects. In 2002, the deflection angle of relativistic, massive particles propagating in a Schwarzschild spacetime was calculated up to the second post-Newtonian order \cite{Accioly2002}.\
In 2014, a strong deflection limit analysis for massive particles in Schwarzschild spacetime was presented for the first time \cite{OYTsupko2014}. Two years later, the same scenario has been considered in Ref.\ \cite{Liu2016}, with more emphasis on possible applications.\ In 2018, the Gauss-Bonnet theorem was applied to derive the deflection angle in the Schwarzschild metric up to second post-Newtonian order \cite{Crisnejo2018}, using a method developed by Gibbons and Werner \cite{Gibbons2008}; see also Ref.\ \cite{Crisnejo-2019}, where this method is extensively applied. In 2019, the deflection angle calculation was extended to the Reissner-Nordström spacetime, considering both weak and strong deflection approximations \cite{Pang2019}.
In 2020, the deflection for massive particles in the Schwarzschild-de Sitter metric was derived up to the second post-Newtonian order \cite{He2020}. Most recently, an analytically exact treatment of the gravitational lensing of massive particles in the charged Newman-Unti-Tamburino metric was presented by Frost \cite{Frost2023}. For a discussion on the black hole shadows of massive particles, the reader may refer to Refs.\ \cite{Kobialko-Bogush-2024,Kobialko-Galtsov-2024}; see also Refs.\ \cite{Kobialko-Bogush-2022, Bogush-Kobialko-2023}. For exact calculations of timelike and null geodesics using various special functions, the reader may consult Refs.\ \cite{Hackmann-2008a, Hackmann-2008b, Cieslik-2022, Cieslik-2023}.

Quite surprisingly, a strong deflection limit analysis for massive particles that applies to any static, asymptotically flat and spherically symmetric spacetime is still missing. In this paper, our goal is to fill this gap, following the procedure outlined in Ref. \cite{Bozza2002}. Specifically, to write the general formulas, we apply the results from Sec. III of Ref. \cite{Feleppa2024}, where a strong deflection limit analysis for photons propagating in a spherically symmetric spacetime surrounded by plasma is discussed. The connection between the two physically different scenarios --- the one discussed here and the other in Ref.\ \cite{Feleppa2024} --- can be understood by noting that a photon wave packet in homogeneous plasma behaves like a massive particle with a velocity equal to the group velocity of the wave packet, a rest mass equal to the plasma frequency and an energy equal to the photon energy, see Refs.\ \cite{Loeb1992, Broderick-2003, BK-Tsupko-2010} for details.

We can consider several scenarios in which our formulas could be applied. One such scenario is the gravitational deflection of massive particles. If a distant source emits massive particles, they can be deflected by a gravitating body, forming multiple images and exhibiting other effects analogous to light lensing. For a detailed discussion, see Frost \cite{Frost2023}. Since the focus here is on neutral particles (where electromagnetic interaction is not involved), neutrinos are the primary candidates for this consideration. For instance, neutrinos emitted by supernovae \cite{Quimby2013} can experience strong deflection when passing near black holes \cite{Mena2007}. The lensing of neutrinos specifically has been studied in works such as \cite{Gerver-1988-neutrino, Escribano-2001-neutrino, Mena2007, Eiroa2008, Luo-2009-neutrino, Adrian-Martinez-2014-neutrino, Coriano-2015-neutrino}. Typically, neutrinos are approximated as massless particles, and lensing effects are calculated using formulas originally derived for light rays. For example, Eiroa and Romero \cite{Eiroa2008} investigated the deflection of neutrinos by very compact objects, applying bending angle formulas intended for light. Although neutrinos travel at velocities very close to the speed of light, the bending angles for massless and massive particles are, in principle, different. 

A second scenario for applying our formulas is in the study of extreme mass ratio inspirals (EMRIs), where a compact object, such as a stellar-mass black hole or neutron star, orbits a supermassive black hole. For a more detailed discussion, the reader is referred to Ref.\ \cite{OYTsupko2014}. EMRIs are sources of gravitational waves and are a primary target for the planned Laser Interferometer Space Antenna (LISA) mission \cite{Amaro-Seoane-2007, Amaro-Seoane-2017, Amaro-Seoane-2018}. Of particular interest in this context are studies of pulsars in close orbits around supermassive black holes \cite{Hackmann-Dhani-2019, Ben-Salem-Hackmann-2022}.

The paper is organized as follows. In Sec.~\ref{sec:exact-deflection}, we consider the unbound motion of a massive test particle in a static, spherically symmetric and asymptotically flat spacetime, and derive the well-known exact expression for the deflection angle of such a particle in terms of its energy per unit rest mass. In Sec.\ \ref{sec:III}, we obtain the equation from which the radial coordinate of unstable circular orbits of massive particles can be deduced. In Sec.\ \ref{sec:IV}, starting from the exact expression for the deflection angle derived in Sec.\ \ref{sec:exact-deflection}, we present a strong deflection limit analysis. The general formulas from Sec.\ \ref{sec:IV} are then applied to three particular cases: the Schwarzschild metric, the Reissner-Nordström metric and finally the Janis-Newman-Winicour metric, see Sec.\ \ref{sec:applications}). We conclude, in Sec.\ \ref{sec:conclusions}, with a summary of the results obtained.

In what follows, we set $G = c = 1$ and use the signature convention $\{-, +, +, +\}$. Moreover, we will work in Schwarzschild coordinates $(t, r, \theta, \varphi)$. Greek indices sum over these coordinates. 

\section{Deflection angle of massive particles in static and spherically symmetric spacetimes}
\label{sec:exact-deflection}

Let us consider an unbound orbit of a massive particle moving from infinity towards a very compact central object, bending its trajectory, reaching a minimum radial coordinate (also referred to as ``closest approach distance''), say $r_{0}$, and then going to infinity again. Restricting our attention to the case of a lensing object described by a static, spherically symmetric and asymptotically flat spacetime, we will now derive an exact expression for the deflection angle of the particle, in integral form. We will follow and generalize the derivation from Refs.\ \cite{Misner1973, Chandrasekhar1983}, where the Schwarzschild metric was considered. See also Refs.\ \cite{Hobson-2006,Weinberg1972}.

The equations governing the geodesics in a spacetime with the line element
\begin{equation}
    ds^2 = g_{\mu\nu} dx^{\mu} dx^{\nu}
\end{equation}
can be derived from the Lagrangian
\begin{equation} \label{Lagrangian}
    \mathcal{L} = \frac{1}{2} \, g_{\mu\nu} \frac{dx^{\mu}}{d\tau} \frac{dx^{\nu}}{d\tau},
\end{equation}
where $\tau$ is some affine parameter along the geodesic. For timelike geodesics, $\tau$ may be identified with the proper time of the particle. 

The line element describing the class of spacetimes under investigation can be written as
\begin{align}
    ds^2 &= g_{\mu\nu} dx^{\mu} dx^{\nu} \nonumber \\
    &= -A(r) \, dt^2 + B(r) \, dr^2 + C(r) \, d\Omega^2,
\end{align}
where $d\Omega^2 \coloneqq d\theta^2 + \sin^2 \theta \, d\varphi^2$ defines the round metric on the unit two-sphere. As anticipated, we assume the spacetime to be asymptotically flat, so the metric coefficients $A(r), B(r)$ and $C(r)$ satisfy
\begin{equation}
    \lim_{r \to \infty} A(r) = 1, \quad
    \lim_{r \to \infty} B(r) = 1, \quad
    \lim_{r \to \infty} C(r) = r^2.
\end{equation}
The metric coefficients are positive in the asymptotically flat region. Approaching the origin, some of them may change sign; we denote the interval in which all the metric coefficients are strictly positive as $\tilde{r} < r < \infty$ (in the case of black holes, such interval extends from the radius of the event horizon to infinity).

The Lagrangian \eqref{Lagrangian}, specialized to our case, becomes
\begin{equation}
    \mathcal{L} = \frac{1}{2}\left[-A(r)\Dot{t}^2 + B(r)\Dot{r}^2 + C(r) \Dot{\varphi}^2\right],
\end{equation}
where, without loss of generality, we considered the orbit of our particle to be confined to the equatorial plane (i.e., we set $\theta = \pi/2$). In the above expression, the dot denotes differentiation with respect to the proper time $\tau$. Now, the equation of motion for $\Dot{t}$ gives
\begin{equation}
    \frac{\partial \mathcal{L}}{\partial t} - \frac{d}{d\tau}\frac{\partial \mathcal{L}}{\partial \Dot{t}} = 0 \hspace{0.2cm} \Rightarrow \hspace{0.2cm} \Dot{t} = \frac{E}{A(r)}, \hspace{0.2cm} E = \text{constant},
\end{equation}
while the one for $\Dot{\varphi}$ leads to
\begin{equation} \label{phidot}
   \frac{\partial \mathcal{L}}{\partial \varphi} - \frac{d}{d\tau}\frac{\partial \mathcal{L}}{\partial \Dot{\varphi}} = 0 \hspace{0.2cm} \Rightarrow \hspace{0.2cm} \Dot{\varphi} = \frac{L}{C(r)}, \hspace{0.2cm} L = \text{constant}.
\end{equation}
In an asymptotically flat spacetime, as the one we are considering, the constant $E$ represents the energy at infinity per unit rest mass of the particle, while the constant $L$ is the angular momentum per unit rest mass of the particle.\ From the condition $2\mathcal{L} = -1$, that is,
\begin{equation}
g_{\mu\nu} \frac{dx^{\mu}}{d\tau} \frac{dx^{\nu}}{d\tau} = -1,
\end{equation}
we can immediately derive an equation for $\Dot{r}$, given by
\begin{equation}
    -A(r)\Dot{t}^2 + B(r)\Dot{r}^2 + C(r)\Dot{\varphi}^2 = -1.
\end{equation}
The above equation can be also rewritten as
\begin{equation} \label{rdot}
    \Dot{r}^2 + \frac{1}{B(r)}\left(1 + \frac{L^2}{C(r)}\right) = \frac{E^2}{A(r)B(r)}.
\end{equation}
By combining Eqs.\ \eqref{phidot} and \eqref{rdot}, the orbit equation of the particle can be easily derived as
\begin{align} \label{orbitequation1}
    \frac{d\varphi}{dr} &= \pm \frac{L}{C(r)}\frac{1}{\sqrt{\dfrac{E^2}{A(r) B(r)} - \dfrac{1}{B(r)}\left(1 + \dfrac{L^2}{C(r)}\right)}} \nonumber \\
    &= \pm \frac{1}{C(r)}\frac{\sqrt{B(r)}}{\sqrt{\dfrac{1}{L^2}\left(\dfrac{E^2}{A(r)} - 1\right) - \dfrac{1}{C(r)}}} \, .
\end{align}
Suppose the test particle moves such that its angular coordinate $\varphi$ increases. In this case, the positive sign in the above equation indicates motion where the coordinate $r$ is also increasing, while the negative sign corresponds to motion where $r$ decreases.

Eq.\ \eqref{orbitequation1} can be compared, for example, with Eq.\ (102) of Ref.\ \cite{Crisnejo2018}. The case $C(r) = r^2$ was also considered in Ref.\ \cite{Weinberg1972}.

By imposing that $dr/d\varphi$ vanishes at $r_{0}$, we obtain a simple relation between $L$ and the distance of closest approach $r_{0}$, namely
\begin{equation} \label{L-E relation}
    L^{2} = C(r_{0})\left(\frac{E^2}{A(r_{0})} - 1\right).
\end{equation}
Unlike the trajectories of light rays, which are determined by a single parameter (typically the closest approach distance or the impact parameter), the trajectories of massive particles in this spacetime depend on two parameters. For the equation \eqref{orbitequation1}, these parameters are $E$ and $L$. Since the closest approach distance $r_{0}$ will appear as an integration limit in the formula for the deflection angle (see below), it is convenient to eliminate another variable, either $E$ or $L$, using Eq.\ \eqref{L-E relation}. This allows us to express the equations in terms of either the pair $(r_{0}, E)$ or the pair $(r_{0}, L)$; see Ref.\ \cite{OYTsupko2014}.

Substituting Eq.\ \eqref{L-E relation} into Eq.\ \eqref{orbitequation1}, we can write
\begin{equation} \label{orbitequation2}
    \frac{d\varphi}{dr} = \pm \sqrt{\frac{B(r)}{C(r)}}\left[\frac{C(r)A(r_0)}{C(r_0)A(r)}\frac{1 - \dfrac{A(r)}{E^2}}{1 - \dfrac{A(r_{0})}{E^2}} - 1 \right]^{-1/2}.
\end{equation}
Above, we present the right-hand side of the orbit equation in terms of $r_0$ and $E$. As discussed earlier, by using Eq.\ \eqref{L-E relation}, we could also express it in terms of $r_0$ and the angular momentum $L$. However, in this article, we will focus exclusively on the $(r_0, E)$ pair for two reasons. First, the energy at infinity, $E$, is directly related to the velocity at infinity, which is a very convenient parameter for applications. Second, expressing formulas in terms of $E$ allows us to take advantage of the equivalence between the motion of massive particles in a vacuum and the motion of photons in a homogeneous plasma discussed in our previous paper \cite{Feleppa2024}. Indeed, it is immediate to notice the mathematical equivalence between Eq.\ (\ref{orbitequation2}) and the radicand of Eq.\ (9) in Ref.\ \cite{Feleppa2024} specialized to the case of homogeneous plasma. More specifically, we notice that the quantity $1 - A(r)/E^{2}$ appearing in Eq.\ \eqref{orbitequation2} can be thought of as the homogeneous plasma refractive index, provided that we identify $E^{-2} \leftrightarrow \omega_{e}^{2}/\omega_{\infty}^{2}$, where $\omega_{e}$ represents the plasma frequency (constant in the homogeneous case) and $\omega_{\infty}$ the photon frequency measured by an observer at infinity. 

As already anticipated in Sec.\ I, all the formulas obtained in Sec.\ IV A of Ref.\ \cite{Feleppa2024} can be applied to the present case, by simply considering the replacement above.

In order to simplify the notation, we will adopt the subscript $0$ when a quantity is evaluated at $r_{0}$. Furthermore, we define
\begin{equation}\label{def F}
    \mathcal{F}^2(r,E) \coloneqq 1 - \frac{A(r)}{E^2}.
\end{equation}
Eq.\ \eqref{orbitequation2} can then be rewritten as
\begin{equation}
    \frac{d\varphi}{dr} = \pm \sqrt{\frac{B(r)}{C(r)}}\left(\frac{C(r)}{C_0}\frac{A_0}{A(r)}\dfrac{\mathcal{F}^2(r, E)}{\mathcal{F}^{2}_{0}} - 1\right)^{-1/2}.
\end{equation}
If we now assume that particles travel in such a way that their angular coordinate $\varphi$ increases, then we must choose the positive sign in Eq.\ \eqref{orbitequation2} when the coordinate $r$ is also increasing; conversely, the negative sign is chosen when the coordinate $r$ is decreasing. Therefore, for a particle coming from infinity towards the gravitational object, and then going to infinity again, the change in the angular coordinate $\varphi$ reads
\begin{equation}
    \Delta \varphi = 2\int_{r_{0}}^{\infty} \frac{\sqrt{B}}{\sqrt{C}\sqrt{\frac{C}{C_{0}}\frac{A_{0}}{A}\frac{\mathcal{F}^2}{\mathcal{F}^{2}_{0}} - 1}}dr,
\end{equation}
where, for simplicity, we omitted the dependencies of the various quantities. If the trajectory were a straight line, $\Delta \varphi$ would equal just $\pi$. Thus, we obtain the deflection angle of the particle with energy at infinity $E$ travelling from infinity towards the compact body, reaching a minimum distance $r_{0}$, and then to infinity again, as
\begin{equation}\label{deflection angle}
    \hat{\alpha}(r_{0}, E) = 2\int_{r_{0}}^{\infty} \frac{\sqrt{B}}{\sqrt{C}\sqrt{\frac{C}{C_{0}}\frac{A_{0}}{A}\frac{\mathcal{F}^2}{\mathcal{F}^{2}_{0}}- 1}}dr - \pi.
\end{equation}
In this formula, the coefficients $A$, $B$, and $C$ depend only on $r$, while the function $\mathcal{F}$, defined in \eqref{def F}, depends on both $r$ and $E$. Additionally, we note that the deflection angle of light rays is recovered in the limit $E \to \infty$. In this case, both $\mathcal{F} \to 1$ and $\mathcal{F}_0 \to 1$. For comparison, see, for example, Refs.\ \cite{Weinberg1972, Virbhadra1998}.

Our goal is to find an analytical approximation of the deflection angle \eqref{deflection angle} in the strong deflection limit. Analogous to the behavior of photons, when $r_0$ approaches the radius of the particle's unstable circular orbit ($r_c$), the particle will orbit the black hole several times before finally escaping. The case $r_0 = r_c$ corresponds to particles that fall directly onto the unstable circular orbit, at which point the deflection angle diverges (see Sec.\ \ref{sec:IV}).

Exploiting the mathematical analogy between photons traveling in homogeneous plasma and massive particles mentioned above, it becomes clear that all the general results presented in Ref.~\cite{Feleppa2024} can be applied to the present case with only minor adjustments.

\section{Unstable circular orbits of massive particles} \label{sec:III}

Before proceeding, let us first derive the equation from which one can deduce the radius of the unstable circular orbit of the massive particle once the metric coefficients are specified. As we will see, such equation will depend on $E$ or $L$ and will be valid only in a certain range of these parameters. To do so, let us go back to Eq.\ \eqref{rdot}, which can be recast in the form
\begin{equation} \label{rdot2}
    \Dot{r}^{2} = \frac{1}{B(r)}\left(\frac{E^2}{A(r)} - \frac{L^2}{C(r)} - 1\right).
\end{equation}
The particle follows a circular motion if, at a fixed radius $r = r_{*}$, the following two conditions are satisfied:
\begin{align}
    \frac{dr}{d\tau}\biggr\rvert_{r = r_{*}} &= 0 \, , \\
    \frac{d^{2}r}{d\tau^{2}}\biggr\rvert_{r = r_{*}} &= 0 \, . \label{secondcondition}
\end{align}
The first of the above conditions immediately leads to
\begin{equation} \label{firstcondition}
    L^2 = C(r_{*}) \left(\frac{E^2}{A(r_{*})} - 1\right).
\end{equation}
Now, differentiating Eq.\ \eqref{rdot2} with respect to $\tau$ and dividing by $\Dot{r}$, gives \cite{OTsupko2015}
\begin{equation}
    2\Ddot{r} = \frac{d}{dr}\left[\frac{1}{B(r)}\left(\frac{E^2}{A(r)} - \frac{L^2}{C(r)} - 1\right)\right].   
\end{equation}
Therefore, Eq.\ \eqref{secondcondition} reduces to the requirement
\begin{equation}
    \frac{d}{dr}\left[\frac{1}{B(r)}\left(\frac{E^2}{A(r)} - \frac{L^2}{C(r)} - 1\right)\right]\biggr\rvert_{r = r_{*}} = 0 \, .
\end{equation}
By using Eq.\ \eqref{firstcondition}, we can write the above expression in terms of $E$, resulting in
\begin{equation} \label{equnstablecircorbitE}
    \frac{C^{\prime}(r_{*})}{C(r_{*})} - \frac{A^{\prime}(r_{*})}{A(r_{*})}\left(1 - \frac{A(r_{*})}{E^2}\right)^{-1} = 0 \, .
\end{equation}
The above expression can be compared, for example, with Eq.\ (30) in Ref.\ \cite{Cardoso-2009}.
Let us assume that Eq.\ \eqref{equnstablecircorbitE} admits at least one positive solution which reduces, in the limit $E \to \infty$, to the radius of the photon sphere. Such a solution corresponds to the radius of the unstable circular orbit of the massive particle and will be valid, as already mentioned, within a certain range of values of $E$.

As a consistency check, let us now consider Eq.\ \eqref{equnstablecircorbitE} in the case of a Schwarzschild black hole. By defining the Schwarzschild radius $r_S = 2M$ ($G = c = 1$) as the unit of measure of distances, with $M$ being the mass of the black hole, the metric coefficients can be written as
\begin{align}
    A(r) &= 1 - \frac{1}{r}, \label{Ar} \\
    B(r) &= \left(1 - \frac{1}{r}\right)^{-1}, \label{Br} \\
    C(r) &= r^2. \label{Cr}
\end{align}
The solution of Eq.\ \eqref{equnstablecircorbitE} is
\begin{equation} \label{solunorb}
    r_{c} = r_{c}(E) = \frac{3E^{2} - 4 + E\sqrt{9E^{2} - 8}}{4\left(E^{2} - 1\right)},
\end{equation}
as expected \cite{OYTsupko2014}. In the limit $E \to \infty$, $r_c$ reduces to $3/2$ (radius of the photon sphere, see Ref.\ \cite{Bozza2002}). Now, by introducing the variable
\begin{equation} \label{definition x}
        x = x(E) \coloneqq \sqrt{1 - \frac{8}{9E^2}} \, ,
\end{equation}
Eq.\ \eqref{solunorb} can be rewritten as \cite{OYTsupko2014}
\begin{equation} \label{rcE}
    r_{c} = r_{c}(x) = 3 \frac{1 + x}{1 + 3x} \, .
\end{equation}
From Eq.\ \eqref{definition x}, we deduce that $E \ge \sqrt{8}/3$, which corresponds to the energy of a particle in the innermost stable circular orbit (in units of $2M$, the radius of the innermost stable circular orbit is equal to 3); according to Eq.\ \eqref{rcE}, $r_{c}$ then varies from $3$ to $3/2$ if the energy $E$ varies from $\sqrt{8}/3$ to infinity. However, as we can easily deduce from Eq.\ \eqref{rdot2}, we must impose that $E \ge 1$ in order to allow the particle to escape to $r = \infty$. Thus, the allowed range of $E$ is $[1, \infty)$.

\section{Deflection angle in the strong deflection limit}
\label{sec:IV}

We are now ready to perform a strong deflection limit analysis on Eq.\ \eqref{deflection angle}. In this section, we extend the procedure originally proposed in Ref.\ \cite{Bozza2002}, which applies to light rays that come close to the photon sphere, to the case of massive particles. 

With decreasing $r_{0}$
(the distance of closest approach), the deflection angle increases, causing the particle to circle the compact object several times before escaping. As $r_{0}$ approaches $r_{c}$, the radius of the particle's unstable circular orbit, the deflection angle approaches infinity.

As already mentioned, we can directly apply all the results from Ref.\ \cite{Feleppa2024} to the present case. Most of the formulas in this section are, in fact, obtained from the ones in \cite{Feleppa2024} by appropriately substituting variables and adjusting notation. Nevertheless, while the core methodology remains unchanged, it is important to provide a brief overview of the procedure’s steps here in order to clarify the results, approximations and assumptions that are used in this paper. This includes addressing any modifications to the original framework that are necessary for the current application, as well as providing detailed explanations of the particular cases and scenarios that are discussed in the following sections.

We introduce the variable $z$ as
\begin{equation}
    z = \frac{A(r) - A_{0}}{1 - A_{0}} \, .
\end{equation}
Consequently, we have
\begin{equation} \label{eq:r-via-z}
    r = A^{-1}\left[A_{0} + \left(1 - A_{0}\right)z\right] \, ,
\end{equation}
where $A^{-1}$ denotes the inverse function and the expression in square brackets is its argument. The deflection angle, Eq.\ \eqref{deflection angle}, can then be rewritten as
\begin{align}
    \hat{\alpha}(r_0, E) &= I(r_0, E) - \pi, \label{alpha} \\
    I(r_0, E) &\coloneqq \int_0^1 R(z,r_0,E) f(z,r_0,E)dz, \label{initialintegral} \\
    R(z,r_0,E) &\coloneqq \frac{2\mathcal{F}_0 \sqrt{ABC_0}(1 - A_0)}{CA^\prime}, \label{R} \\
    f(z,r_0,E) &\coloneqq \frac{1}{\sqrt{A_0 \mathcal{F}^2 - [(1 - A_0)z + A_0]\frac{C_0}{C}\mathcal{F}_0^2}}.\label{f}
\end{align}
According to Eq.\ \eqref{eq:r-via-z}, any quantity without the subscript $0$ is evaluated at $r$. In line with Refs.\ \cite{Bozza2002,Feleppa2024}, we approximate $f(z,r_{0},E)$ to second order in $z$, and define
\begin{equation}\label{f_0}
    f_0(z,r_0,E) \coloneqq \frac{1}{\sqrt{\alpha \, z + \beta \, z^2}},
\end{equation}
with $\alpha = \alpha(r_0,E)$ given by
\begin{equation}\label{alphar0}
    \alpha \coloneqq \frac{\mathcal{F}_0^2 \left(1-A_0\right)}{C_0 A_0^\prime}\left[C_0^\prime A_0 + C_0 \left(2A_0 \frac{\mathcal{F}_0^\prime}{\mathcal{F}_0} -A_0^\prime\right)\right],
\end{equation}
and $\beta = \beta(r_0,E)$ expressed as
\begin{multline}\label{betar0}
    \beta \coloneqq \frac{\mathcal{F}_0^2 (1 - A_0)^2}{2C_0^2 A_0^{\prime 3}}\left[2C_0 C_0^\prime A_0^{\prime 2} \right.
    \\
    \left.
    \hspace{-1cm} + \hspace{1mm} A_0^\prime A_0 \left(C_0 C_0^{\prime \prime} - 2C_0^{\prime 2}\right) \right.
    \\
    \left.
    \hspace{0.4cm} - \hspace{1mm} C_0^2 A_0^{\prime \prime} A_0 \left(\frac{C_0^\prime}{C_0}  + 2\frac{\mathcal{F}_0^\prime}{\mathcal{F}_0}\right)\right]
    \\
    + \frac{A_0(1 - A_0)^2}{A_0^{\prime 2}}\left(\mathcal{F}_0^{\prime 2} + \mathcal{F}_0 \mathcal{F}_0^{\prime \prime}\right).
\end{multline}
To examine paths that approach $r_{c}$, we define a parameter $\delta \ll 1$ as follows:
\begin{equation} \label{deltar0}
    r_{0} \coloneqq r_{c} (1 + \delta).
\end{equation}
Note that $r_{c}$ depends on $E$, see Eqs.\ \eqref{rcE} and \eqref{definition x}.

We also notice that, setting $\alpha = 0$, gives the radius of the particle's unstable circular orbit. Indeed, we have
\begin{equation} \label{alpha zero eq}
    \alpha = 0 \hspace{0.1cm} \Rightarrow \hspace{0.1cm} C_0^\prime A_0 - C_0 A_0^\prime = -2C_0 A_0 \frac{\mathcal{F}_0^\prime}{\mathcal{F}_0}.
\end{equation}
Now, recalling the definition of $\mathcal{F}$, Eq.\ \eqref{def F}, we obtain
\begin{equation} \label{radiusunstableorbit}
    \frac{\mathcal{F}_0^\prime}{\mathcal{F}_0} = -\frac{A_0^\prime}{2\left(E^2 - A_0\right)},
\end{equation}
which, substituted into Eq.\ \eqref{alpha zero eq}, leads to
\begin{equation}\label{equco}
   \frac{C_{0}^{\prime}}{C_{0}} - \frac{A_{0}^{\prime}}{A_{0}}\left(1 - \frac{A_{0}}{E^{2}}\right)^{-1} = 0.
\end{equation}
Eq.\ \eqref{equco} coincides with Eq.\ \eqref{equnstablecircorbitE} derived in Sec.\ III.

By introducing the function
\begin{multline}\label{definition g}
    g(z,r_0,E) \coloneqq R(z,r_0,E)f(z,r_0,E) \\
    - R(0,r_c,E)f_0 (z,r_0,E),
\end{multline}
Eq.\ \eqref{initialintegral} can be split into two parts, namely
\begin{align}
    I(r_{0}, E) &= I_{D}(r_{0},E) + I_{R}(r_{0},E), \\
    I_{D}(r_{0},E) &\coloneqq \int_0^1 R(0,r_c,E)f_0 (z,r_0,E)dz, \\
    I_{R}(r_{0},E) &\coloneqq \int_0^1 g(z,r_0,E)dz. \label{IR}
\end{align}
The subscript $D$ stands for ``divergent'' (since $I_{D}$ indeed diverges in the limit $r_{0} \to r_{c}$), while the subscript $R$ stands for ``regular''. By explicitly calculating $I_{D}(r_{0},E)$ and subsequently expanding the coefficient $\alpha$ up to first order in $\delta$, we obtain
\begin{equation}
    I_{D}(r_0,E) \simeq -a\log\delta(r_0) + b_D + \mathcal{O}(\delta),
\end{equation}
where we define
\begin{align}
    a &\coloneqq \frac{R(0,r_c,E)}{\sqrt{\beta_c}}, \label{a} \\
    b_D &\coloneqq a\log\frac{2\left(1 - A_c\right)}{A_c^\prime r_c}. \label{b_D}
\end{align}
The quantity $\beta_{c}$ in the above expressions reads
\begin{multline}\label{betac}
    \beta_c = \frac{\mathcal{F}_c\left(1-A_c\right)^2}{2C_c A_c^{\prime 2}}\left[\mathcal{F}_c \left(C_c^{\prime \prime}A_c - C_c A_c^{\prime \prime}\right)\right.
    \\
    \left. + \left(3C_c^\prime A_c + C_c A_c^\prime \right)\mathcal{F}_c^\prime + 2A_c C_c \mathcal{F}_c^{\prime \prime}\right].
\end{multline}
Note that we use the subscript $c$ to indicate evaluation at $r_c$. Additionally, we remind the reader that the function $R(z,r_c,E)$ is defined in Eq.\ \eqref{R}, while $r_c$ is given by Eq.\ \eqref{rcE}. Now, expressing Eq.\ \eqref{IR} as a series in $\delta$ and retaining only the leading term gives
\begin{equation}\label{regular term general}
    I(r_{0},E) = \int_{0}^{1}g(z,r_{c},E)dz + \mathcal{O}\left(\delta\right) \coloneqq b_{R}.
\end{equation}
In the strong deflection limit, the deflection angle \eqref{alpha} is then found to be
\begin{equation} \label{deflection angle final}
    \hat{\alpha}(r_0,E) = -a\log\delta(r_0,E) + b,
\end{equation}
Above, $\delta(r_0,E) = r_0/r_{c}(E) - 1 \ll 1$, see Eq.\ \eqref{deltar0}, $a$ is defined by Eq.\ \eqref{a} and $b \coloneqq b_D + b_R - \pi$, with $b_D$ and $b_R$ given by Eqs.\ \eqref{b_D} and \eqref{regular term general}, respectively.

Eq.\ \eqref{deflection angle final} can also be expressed as a function of the impact parameter $u$. For massive particles, the impact parameter is defined as \cite{Misner1973}
\begin{equation}
    u^{2} = \frac{L^{2}}{E^{2} - 1}.
\end{equation}
Using Eq.\ \eqref{L-E relation}, we can express $u$ in terms of the closest approach distance $r_{0}$ and the particle's energy $E$ as
\begin{equation} \label{impactpar}
    u^{2} = \frac{\mathcal{F}_{0}^{2}}{1 - \frac{1}{E^{2}}}\frac{C_{0}}{A_{0}}.
\end{equation}
In the limit $r_{0} \to r_{c}$, $u$ converges to $u_{c}$, which corresponds to the value for massive particles coming from infinity and entering the circular orbit.
Therefore, we introduce a parameter $\varepsilon \ll 1$ through the equation
\begin{equation} \label{epsilon}
    u \coloneqq u_{c} (1 + \varepsilon), \quad u_{c} = \sqrt{\frac{\mathcal{F}_{c}^{2}}{1 - \frac{1}{E^{2}}}\frac{C_{c}}{A_{c}}}.
\end{equation}
Next, by approximating $u$ for small $\delta$, it follows that
\begin{equation}
    u - u_{c} = \ell \, r_{c}^2 \delta^2 = u_{c} \varepsilon,
\end{equation}
with $\ell$ given by
\begin{equation}
    \ell \coloneqq \frac{\beta_{c} A_{c}^{\prime 2}\sqrt{C_{c}}}{2 \mathcal{F}_{c} A_{c}^{3/2}(1 - A_{c})^2 \sqrt{1 - 1/E^{2}}}.
\end{equation}
In the strong deflection limit, as a function of the impact parameter, Eq.\ \eqref{deflection angle final} results in
\begin{equation}\label{deflection angle imppar}
    \hat{\alpha}(u,E) = -\Bar{a}\log\varepsilon(u,E) + \Bar{b},
\end{equation}
with $\varepsilon(u,E) = u/u_{c}(E) - 1 \ll 1$ and the coefficients $\Bar{a}$ and $\Bar{b}$ expressed as
\begin{align}
    \Bar{a} &\coloneqq \frac{a}{2} = \frac{R(0,r_{c},E)}{2\sqrt{\beta_{c}}}, \label{abarimp} \\
    \Bar{b} &\coloneqq -\pi + b_R + \Bar{a}\log\frac{2\beta_{c}}{\mathcal{F}_{c}^2 A_{c}}, \label{bbarSDLgeneral}
\end{align}
respectively.

As anticipated, in certain applications, it may be more convenient to use the asymptotic velocity at infinity, denoted by $v$, instead of the energy $E$. To rewrite all the formulas in terms of velocity rather than energy, we can simply perform the substitution
\begin{equation}\label{eq:E-velocity} 
    E = \frac{1}{\sqrt{1-v^2}}.
\end{equation}

\section{Applications} 
\label{sec:applications}

The general formulas presented in the previous section, valid for static, asymptotically flat and spherically symmetric spacetimes, can now be specialized to particular cases. After reproducing the already known results for the Schwarzschild metric (see Ref.\ \cite{OYTsupko2014} for further details), we will consider the Reissner-Nordström and Janis-Newman-Winicour metrics.

\subsection{Schwarzschild metric}\label{SCHW}

As already anticipated in Sec.\ \ref{sec:III}, in this case the metric coefficients are
\begin{equation}
    A(r) = 1 - \frac{1}{r}, \quad
    B(r) = \left(1 - \frac{1}{r}\right)^{-1}, \quad C(r) = r^2.
\end{equation}
The two functions $R(z,r_{0},E)$ and $f(z,r_{0},E)$ defined in Eqs.\ \eqref{R} and \eqref{f}, respectively, read
\begin{align}
    R(z,r_{0},E) &= 2\mathcal{F}_{0} = 2\sqrt{1 - \left(1 -\frac{1}{r_{0}}\right)\frac{1}{E^{2}}}, \\
    f(z,r_{0},E) &= \frac{1}{\sqrt{\alpha z + \beta z^2 - \gamma z^3}}, \label{fmp}
\end{align}
where the coefficients $\alpha, \beta$ and $\gamma$ are given by
\begin{align}
    \alpha = \alpha(r_{0},E) &= 2 - \frac{3}{r_0} - \frac{2}{E^{2}} \left(1 - \frac{2}{r_0} + \frac{1}{r_0^2}\right), \label{alpha SCHW} \\
    \beta = \beta(r_{0},E) &= \frac{3}{r_0} - 1 + \frac{1}{E^{2}} \left(1 - \frac{4}{r_0} + \frac{3}{r_0^2}\right), \label{beta SCHW} \\
    \gamma = \gamma(r_{0},E) &= \frac{1}{r_0}\left[1 - \frac{1}{E^{2}} \left(1 - \frac{1}{r_0}\right)\right].
\end{align}
As $E \to \infty$, it is immediate to verify that Eq.\ \eqref{fmp} reduce to Eq.\ (42) of Ref.\ \cite{Bozza2002}. Now, the radius of the unstable circular orbit can be found by setting $\alpha(r_{0},E) = 0$, resulting in Eq.\ \eqref{rcE}, with $x = x(E)$ defined in Eq.\ \eqref{definition x}. From Eqs.\ \eqref{a}, \eqref{b_D} and \eqref{regular term general}, we then obtain the various coefficients of the strong deflection limit analysis in terms of $x$, that is 
\begin{align}
    a &= a(x) = 2\sqrt{\frac{1 + x}{2x}}, \label{aSchw} \\
    b_D &= b_D (x) = a\log 2, \\
    b_R &= b_R (x) = -a\log\left[\frac{\left(\sqrt{3x - 1} + \sqrt{6x}\right)^2}{24x}\right]. \label{bRSchw}
\end{align}
In terms of the energy per unit rest mass of the particle, as well as of the distance of closest approach, the deflection angle can then be written as
\begin{equation}\label{deflection angle mp}
   \hat{\alpha}(r_0, E) = -2\sqrt{\frac{1 + x}{2x}}\log\left(z_1 (x) \delta(r_{0},x)\right) - \pi,
\end{equation}
where the quantity $z_1$ is defined by
\begin{equation}
    z_1 (x) \coloneqq \frac{9x - 1 + 2\sqrt{6x(3x - 1)}}{48x} \, ,
\end{equation}
with $x = x(E)$ given by Eq.\ \eqref{definition x}. Eq.\ \eqref{deflection angle mp} agrees with Eq.\ (53) in Ref.\ \cite{OYTsupko2014}. 
\begin{figure}[t]
\includegraphics[width=8cm]{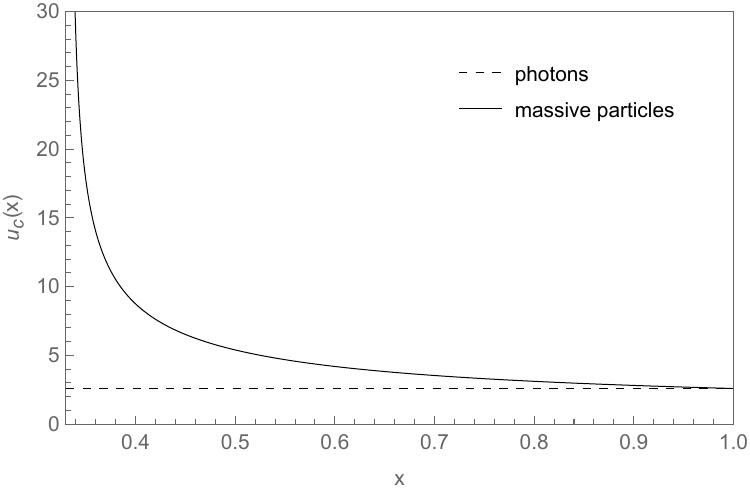}
\centering
\caption{Critical impact parameter for massive particles given by Eq.\ \eqref{minimum impact parameter} as a function of the variable $x = x(E)$, which is defined in Eq.\ \eqref{definition x}. In the limit $x \to 1$ (which corresponds to the limit $E \to \infty$), the critical impact parameter approaches the corresponding value for photons, i.e., $\lim_{x \to 1} \Bar{u}_{c}(x) = 3\sqrt{3}/2$ \cite{Bozza2002}. In the plot, this is represented by the dotted horizontal line. The critical impact parameter for deflection increases for lower-energy particles, diverging for particles at rest at infinity, corresponding to $x = 1/3$.}
\end{figure}
\begin{figure}[t]
\includegraphics[width=8.1cm]{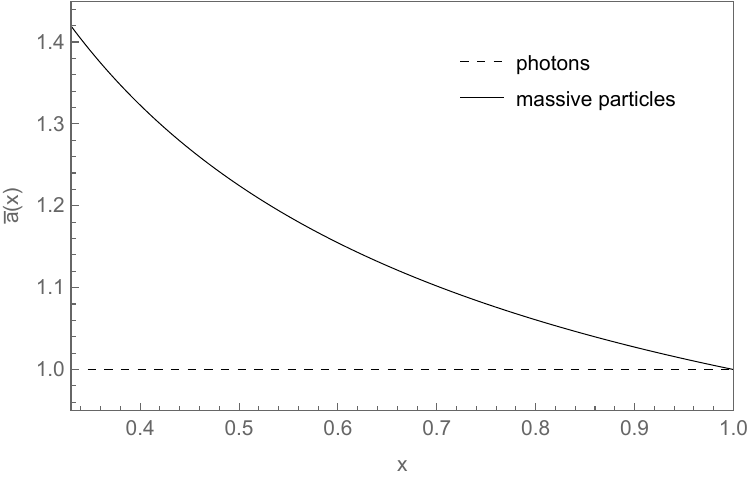}
\centering
\caption{Behavior of the strong deflection limit coefficient $\Bar{a}(x)$ defined in Eq.\ \eqref{abarSCHW} in Sec.\ \ref{SCHW}, which represents the coefficient of the logarithmic divergence. As expected, in the limit $x \to 1$, $\Bar{a}(x)$ approaches the corresponding value for photons, i.e., $\lim_{x \to 1} \Bar{a}(x) = 1$ \cite{Bozza2002}.}
\end{figure}
\begin{figure}[t]
\includegraphics[width=8.2cm]{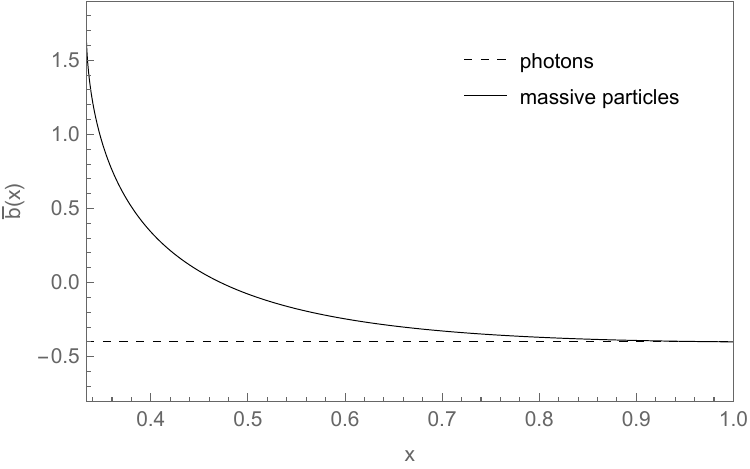}
\centering
\caption{Behavior of the strong deflection limit coefficient $\Bar{b}(x)$ defined in Eq.\ \eqref{bbarSCHW}, Sec.\ \ref{SCHW}.\ As expected, in the limit $x \to 1$, $\Bar{a}(x)$ approaches the corresponding value for photons, i.e., $\lim_{x \to 1} \Bar{b}(x) = -0.4002$ \cite{Bozza2002}.}
\end{figure}
\begin{figure}[!h]
\includegraphics[width=8.4cm]{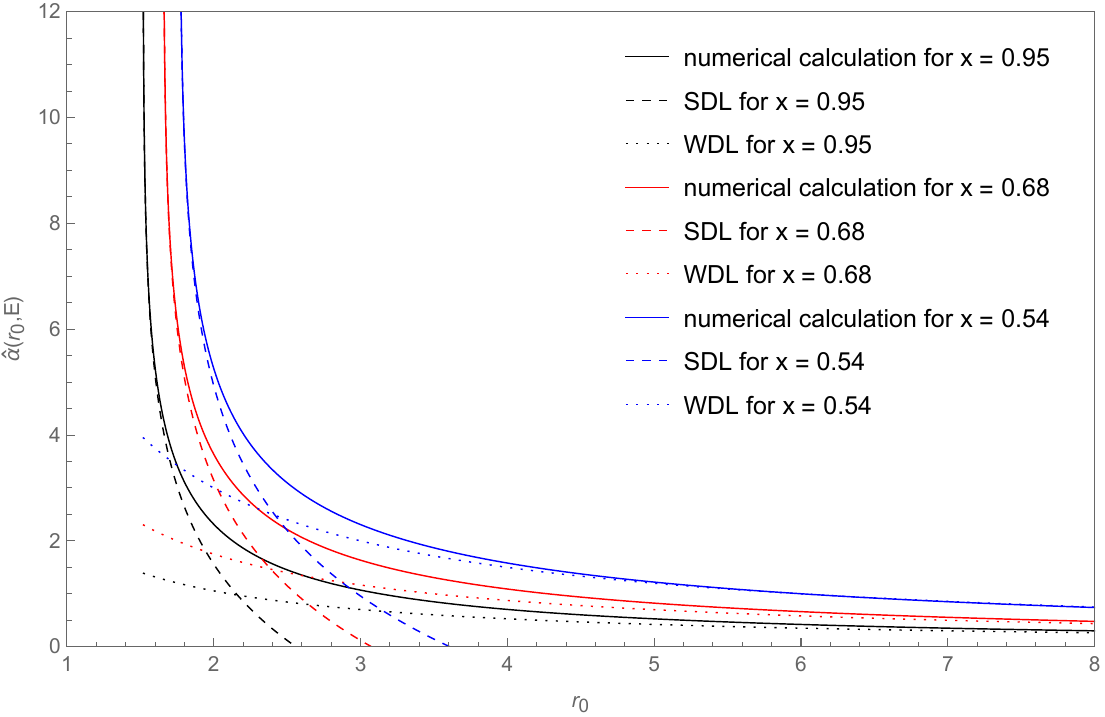}
\centering
\caption{A comparison of the deflection angle for massive particles in a Schwarzschild metric obtained numerically starting from Eq.\ \eqref{deflection angle}, with that in the weak and strong deflection limits (WDL and SDL, respectively), for different values of $x$. The deflection angle in a Schwarzschild spacetime in the strong deflection limit is given by Eq.\ \eqref{deflection angle mp}, while the formula valid in the weak deflection limit can be found in Ref.\ \cite{Crisnejo2018}, Eq.\ (109). Note that Eq.\ \eqref{deflection angle}, and Eq.\ (109) in Ref.\ \cite{Crisnejo2018}, are expressed in terms of $E$ and $v$, respectively. Using Eqs.\ \eqref{definition x} and \eqref{eq:E-velocity}, we can easily express them in terms of $x$.}
\end{figure}

In terms of the impact parameter $u$, the deflection angle instead results in
\begin{equation}\label{daimp}
    \hat{\alpha}(u, E) = -\Bar{a}(x)\log\varepsilon(u,x) + \Bar{b}(x),
\end{equation}
where $\varepsilon(u,x)$, $\Bar{a}(x)$ and $\Bar{b}(x)$ read
\begin{align}
    \varepsilon(u,x) &= \frac{u}{u_{c}(x)} - 1,\\
    \Bar{a}(x) &= \sqrt{\frac{1 + x}{2x}}, \label{abarSCHW} \\
    \Bar{b}(x) &= -\Bar{a}(x)\log\left(\frac{2z_1^2 (x)}{3x}\right) - \pi, \label{bbarSCHW}
\end{align}
respectively, with $u_{c}(x)$ given by
\begin{equation} \label{minimum impact parameter}
   u_{c}(x) = \sqrt{\frac{3(1 + x)}{3x - 1}}r_{c}(x). 
\end{equation}
Above, $r_{c}(x)$ is given by Eq.\ \eqref{rcE}. Eq.\ \eqref{daimp} is in agreement with Eq.\ (65) in Ref.\ \cite{OYTsupko2014}. 
The quantities $u_{c}(x)$, $\Bar{a}(x)$ and $\Bar{b}(x)$ are plotted in Figs.\ 1, 2 and 3, respectively. We note that $u_{c}(x)$ increases as the particle’s energy decreases, diverging when the particle originates from rest at infinity ($x \to 1/3$).\ Both coefficients $\Bar{a}(x)$ and $\Bar{b}(x)$ also increase at lower energies. Thus, a massive particle experiences stronger deflection than a massless particle coming from infinity with the same impact parameter. In Fig.\ 4, we present a comparison between the deflection angle obtained numerically and the one obtained analytically in the weak and strong deflection limits, for different values of $x$. We observe that the deflection angle increases as the energy decreases.

The critical impact parameter for massive particles in the Schwarzschild metric can also be found in Mielnik and Plebański \cite{Mielnik1962} and Zakharov \cite{Zakharov-1994}.

\subsection{Reissner-Nordström metric}\label{RN}

The Reissner-Nordström metric is a static solution to the Einstein--Maxwell field equations, which describe the gravitational field of a charged, nonrotating and spherically symmetric massive object \cite{Reissner1916, Nordström1918}. In units of $2M$, the metric coefficients in this case are
\begin{align}
    A(r) &= B(r)^{-1} = 1 - \frac{1}{r} + \frac{q^{2}}{r^{2}}, \label{Archarged} \\
    C(r) &= r^2, \label{Crcharged}
\end{align}
where $q$ denotes the charge parameter of the black hole. The Reissner-Nordström spacetime possesses two concentric horizons located at
\begin{equation}
    r_{\pm} = \frac{1}{2}\left(1 \pm \sqrt{1 - 4q^{2}}\right).
\end{equation}
As we can deduce from the above equation, these two horizons exist only if $q \in [0, 1/2]$ ($q = 1/2$ corresponds to an extremal black hole), which is thus considered the allowed range for the charge. Therefore, from now on we restrict the charge parameter $q$ to the range $[0, 1/2]$. 

We also note that in the Reissner-Nordström case, when finding the inverse function $A^{-1}$ [see Eq.\ \eqref{eq:r-via-z}], we obtain a quadratic equation and must choose the outermost branch.

Let us proceed by computing the coefficients $\alpha$ and $\beta$, which can easily be found from Eqs.\ \eqref{alphar0} and \eqref{betar0}, respectively. The result is
\begin{widetext}
\begin{align}
    \alpha(r_0) &= \frac{\left(q^2 - r_0 \right) \left\{E^2 r_0^2 \left[r_0 \left(2r_0 - 3\right) + 4q^2 \right] - 2\left[r_0 \left(r_0 - 1\right) + q^2 \right]^2\right\}}{r_0^4 E^2 \left(2q^2 - r_0\right)}, \\
    \beta(r_0) &= \frac{\left(q^2-r_0 \right)^2 \left[r_0^2 \left(r_0 - 3\right) + 9q^2 r_0 - 8q^2 \right] \left[r_0 \left(E^2 r_0 - r_0 + 1\right) - q^2 \right]}{E^2 r_0^4 \left(2q^2 -r_0 \right)^3}.
\end{align}
\end{widetext}
The above relations reduce to Eqs.\ \eqref{alpha SCHW} and \eqref{beta SCHW} in the limit $q \to 0$, and to Eqs.\ (54) and (55) in Ref.\ \cite{Bozza2002} in the limit $E \to \infty$, as expected. In terms of $E$, the radius of the particle's unstable circular orbit can be found by setting $\alpha(r_0) = 0$, resulting in a quartic equation. In the following, we will consider a perturbative approach. Specifically, we will calculate the coefficients $a$, $b_{D}$ and $b_{R}$ up to order $q^2$. Extending the analysis to include higher-order terms is straightforward. For the sake of simplicity, even though perturbative results will be obtained, we will continue using equalities in our equations. The procedure we follow in order to obtain the quadratic correction to the radius of the unstable circular orbit of the massive particle, $r_{c}$, is the same as the one outlined in Sec.\ V of Ref.\ \cite{Perlick2015} for light rays propagating in a plasma environment. In terms of the variable $x = x(E)$, the result one obtains is
\begin{equation}
    r_{c} = r_{c}(x,q) = 3\frac{1 + x}{1 + 3x} - \left(1 + \frac{1}{3x}\right)q^2.
\end{equation}
\begin{figure}
\includegraphics[width=8.4cm]{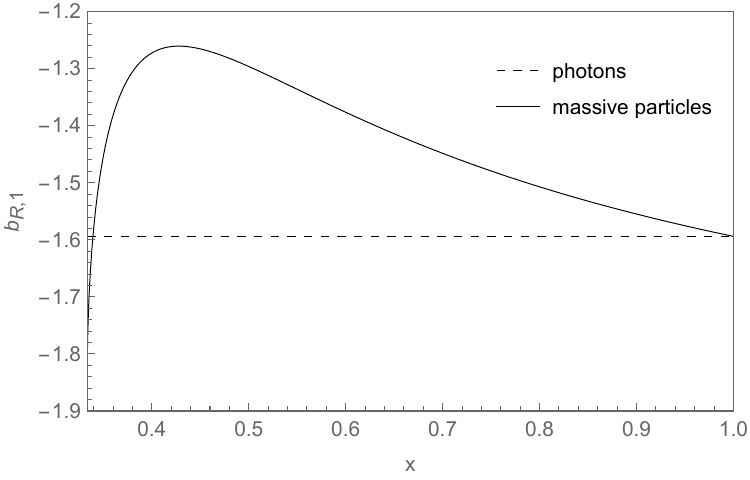}
\centering
\caption{Coefficient $b_{R,1}$ in Eq.\ \eqref{coefficient bR} in Sec.\ \ref{RN}.}
\end{figure}

We can now easily compute the strong deflection limit coefficients $a = a(x,q), b_{D} = b_{D}(x,q)$ and $b_{R} = b_{R}(x,q)$. As for the first two coefficients, from Eqs.\ \eqref{a} and \eqref{b_D}, we obtain
\begin{align}
    a &= \frac{36 x^2 (x + 1) + \left[3x^2 (6x + 5) - 1\right]q^2}{18\sqrt{2x^5 (x+1)}}, \label{coefficient a charged} \\
    b_D &= \frac{1}{18\sqrt{2x^5 (x + 1)}}\left\{36x^2 (x + 1) \log 2 \right. \nonumber \\
    &\hspace{-0.2cm}+ \left. \left\{3x^2 \left[6x (2+\log 2) + 4 + \log 32\right] - \log 2 \right\}q^2 \right\},
    \label{coefficient bD charged}
\end{align}
respectively.

As for the quantity $b_{R}$, we recall that it is defined in terms of an integral. The integration can be performed by expanding the integrand (i.e., $g(z,r_{c},E)$, see Eq.\ \eqref{definition g} in Sec.\ \ref{sec:IV} for the definition of this function) in powers of $q$ and then evaluating the single coefficients \cite{Bozza2002}. Up to order $q^2$, we get
\begin{align}\label{coefficient bR}
    b_{R} = b_{R}(x,q) &= \int_{0}^{1}g(z,r_{c},E)dz \nonumber\\
    &= b_{R,0}(x) + b_{R,1}(x)q^{2},
\end{align}
where $b_{R,0}(x)$ is expressed as
\begin{align}
    b_{R,0}(x) &= -a(x,0) \log\left[\frac{\left(\sqrt{3x - 1} + \sqrt{6x}\right)^2}{24x}\right], \label{coeff uncharged} \\
    a(x,0) &= 2\sqrt{\frac{1 + x}{2x}}, \label{ax0}
\end{align}
i.e., the result for an uncharged black hole (Eqs.\ \eqref{coeff uncharged} and \eqref{ax0} indeed coincide with Eqs.\ \eqref{bRSchw} and \eqref{aSchw}, respectively), while the coefficient $b_{R,1}(x)$ is defined as
\begin{equation}\label{br1q}
    b_{R,1} (x) \coloneqq \int_0^1 \mathrm{b}_{R,1}(z,x)dz,
\end{equation}
where, in turn, $\mathrm{b}_{R,1}(z,x)$ reads
\begin{widetext}
\begin{multline}
    \mathrm{b}_{R,1}(z,x) \coloneqq \frac{(1 + 3x)^{3/2}}{36 x^2}\left(\frac{\sqrt{2}\left[1 - 3x (1 + 2x)\right]}{z\sqrt{x (1 + x)(1 + 3x)}}\right. \\
    \left. + \hspace{0.5mm} \frac{4 \sqrt{3} x(1 + x)(1 + 3x)z^2 \left\{3x \left[6 x(z^2 - 3z + 1) + z(2z - 1) + 3\right] + z - 3\right\}}{\left\{(1 + x)(1 + 3x)\left[3x
   (2 - z) + z\right]z^2 \right\}^{3/2}}\right).
\end{multline}
\end{widetext}
The result of the integral in Eq.\ \eqref{coefficient bR} is shown in Fig.\ 5.
As we can observe, for small values of $x$ (i.e., for small $E$), the quadratic correction deviates significantly from the corresponding value calculated for photons \cite{Bozza2002}.

The deflection angle can then be written as a function of $r_0$, as well as of $x$ and $q$, as
\begin{multline} \label{deflanglechargedr0}
    \hat{\alpha}(r_{0},E,q)  = -a(x,q)\log\delta(r_{0},x,q) \\
    + b_{D}(x,q) + b_{R}(x,q) - \pi,
\end{multline}
where $a(x,q), b_{D}(x,q)$ and $b_{R}(x,q)$ are given by Eqs.\ \eqref{coefficient a charged}, \eqref{coefficient bD charged} and \eqref{coefficient bR}, respectively. We also recall that the quantity $\delta(r_{0},x,q)$ is defined by Eq.\ \eqref{deltar0}. The formula we derived for the deflection angle, namely Eq.\ \eqref{deflanglechargedr0}, is valid up to order $q^2$.

In terms of the impact parameter $u$, using Eqs.\ \eqref{abarimp} and \eqref{bbarSDLgeneral}, we obtain
\begin{equation} \label{deflanglecharged}
    \hat{\alpha}(u,E,q)   = -\Bar{a}(x,q)\log\varepsilon(u,x,q) + \Bar{b}(x,q),
\end{equation}
where the coefficient $\Bar{a}(x,q)$ is simply given by $a(x,q)/2$, with $a(x,q)$ expressed by Eq.\ \eqref{a}, while $\Bar{b}(x,q)$ is calculated from Eq.\ \eqref{bbarSDLgeneral}, resulting in
\begin{widetext}
    \begin{equation} \label{bbar charged}
        \Bar{b}(x,q) = -\pi + b_{R}(x,q) + \frac{2 q^2 \left(27 x^3 + 3 x^2 + x + 1\right)+\left\{q^2 \left[3x^2 (6 x+5) - 1\right] + 36 x^2 (x + 1)\right\} \log (6 x)}{36 \sqrt{2x^5 (x+1)}}.
    \end{equation}
\end{widetext}
In the expression for the deflection angle, Eq.\ \eqref{deflanglecharged}, the quantity $\varepsilon(u,x,q)$ reads
\begin{equation}
    \varepsilon(u,x,q) = \frac{u}{u_{c}(x,q)} - 1,
\end{equation}
where the critical impact parameter $u_{c}(x,q)$ is obtained from Eq.\ \eqref{impactpar} evaluated at $r = r_{c}$, giving
\begin{equation} \label{ipcharged}
    u_{c}(x,q) = \frac{3 - q^2 - 3x(q^2 - 1)}{1+ 3x}\sqrt{1 + \frac{4}{3x - 1}}.
\end{equation}
The critical impact parameter for massive particles in the Reissner-Nordström metric was also carefully investigated in Ref.\ \cite{Zakharov-1994}.

As anticipated in Sec.\ I, the deflection of massive particles was also studied in Ref. \cite{Pang2019}, where an expression in terms of elliptic integrals was derived. The authors considered various limits but did not derive an expansion in the strong deflection limit comparable to ours. Their Eq.\ (50) was obtained by expanding the deflection angle around the Schwarzschild photon sphere radius rather than the Reissner-Nordström one. Such an expansion cannot describe the higher-order images in the Reissner-Nordström case, as it does not converge in the strong deflection regime. Moreover, in their treatment, they considered an exact solution for the unstable circular orbit radius, whereas we retained only the quadratic correction in the charge.

We conclude by showing, in Figs.\ 6, 7 and 8, the quantities $u_{c}(x,0) - u_{c}(x,q)$, $\Bar{a}(x,0) - \Bar{a}(x,q)$ and $\Bar{b}(x,0) - \Bar{b}(x,q)$, respectively, as functions of $x$ for different values of $q$. The differences between the various curves vary with the energy. Moreover, for a fixed $q$, $u_{c}(x,q)$ decreases as $q$ increases (a consequence of the fact that the horizon shrinks), while $\Bar{a}(x,q)$ and $\Bar{b}(x,q)$ increases as $q$ increases.
\begin{figure}[!t]
\includegraphics[width=8.3cm]{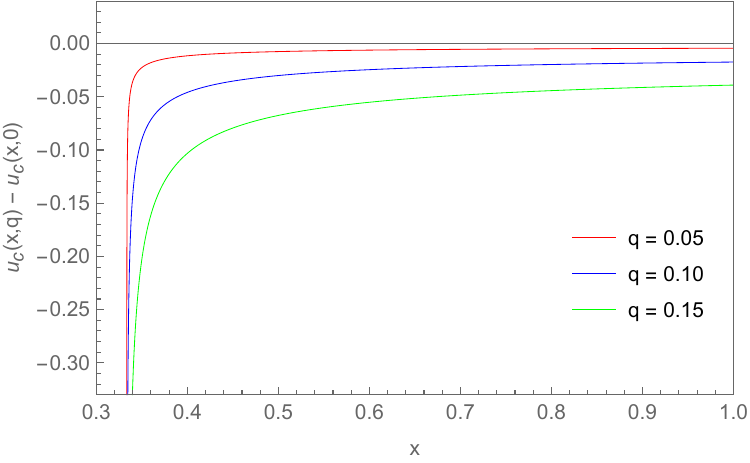}
\centering
\caption{Difference $u_c (x,q) - u_c (x,0)$ as a function of the particle's energy at infinity, for different values of the charge parameter $q$. The critical impact parameter $u_c (x,q)$ is defined in Eq.\ \eqref{ipcharged}, Sec.\ \ref{RN}.}
\end{figure}
\begin{figure}[!t]
\includegraphics[width=8.3cm]{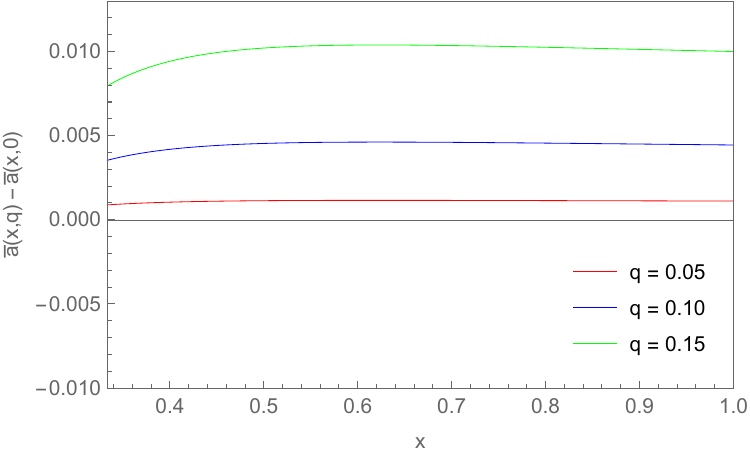}
\centering
\caption{Difference $\Bar{a}(x,q) - \Bar{a}(x,0)$ as a function of the particle's energy, for different values of the charge parameter $q$.\ The strong deflection limit coefficient $\Bar{a}(x,q)$ is defined by $a(x,q)/2$, with $a(x,q)$ given by Eq.\ \eqref{coefficient a charged} (Sec.\ \ref{RN}).}
\end{figure}
\begin{figure}[!t]
\includegraphics[width=8.35cm]{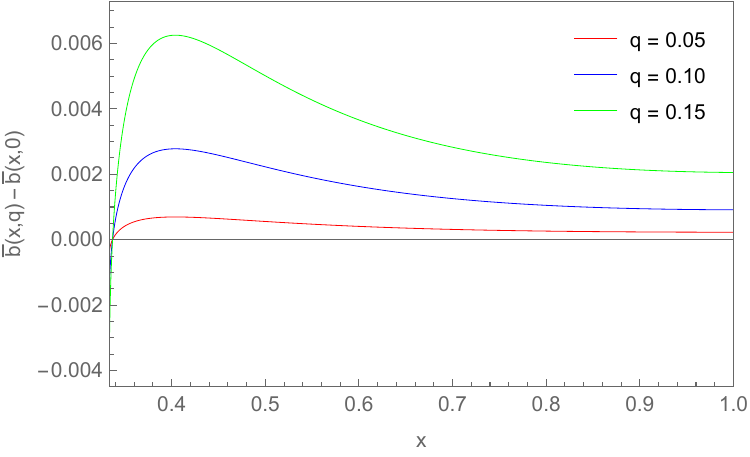}
\centering
\caption{Difference $\Bar{b}(x,q) - \Bar{b}(x,0)$ as a function of the particle's energy, for different values of charge parameter $q$. The strong deflection limit coefficient $\Bar{b}(x,q)$ is given by Eq.\ \eqref{bbar charged} in Sec.\ \ref{RN}.}
\end{figure}

\subsection{Janis-Newman-Winicour metric}\label{JNW}

Here, we consider the Janis-Newman-Winicour metric, which is the unique static and spherically symmetric solution to the Einstein equations for a massless, minimally coupled scalar field \cite{Fisher1948,JNW1968,Wyman1981,Virbhadra1997}. This solution is characterized by the mass $M$ and a parameter $\gamma$ related to the scalar field strength. Taking $r_{\gamma} = 2M/\gamma$ as the unit of measure of distances, the metric coefficients are \cite{Aratore-Tsupko-Perlick-2024}
\begin{align}
    A(r) &= \left(1 - \frac{1}{r}\right)^{\gamma}, \label{AJNW} \\
    B(r) &= \left(1 - \frac{1}{r}\right)^{-\gamma}, \label{BJNW} \\
    C(r) &= \left(1 - \frac{1}{r}\right)^{1 - \gamma} r^2, \label{CJNW}
\end{align}
with the scalar field given by
\begin{equation}
    \Phi(r) = \sqrt{\frac{1 - \gamma^2}{16\pi}}\log\left(1 - \frac{1}{r}\right).
\end{equation}
The dimensionless parameter $\gamma$ appearing in the above relations takes values in the range $0 < \gamma \le 1$. However, as we shall see, the perturbative character of the analysis below significantly narrows down the admissible range of $\gamma$. As one can easily check, when $\gamma = 1$, the Schwarzschild solution is recovered. Moreover, as shown in Ref.\ \cite{Joshi1997}, the Janis-Newman-Winicour solution has a naked curvature singularity located at $r = r_\gamma$. When $\gamma$ is relatively close to unity, the spacetime under investigation admits the existence of unstable circular orbits which are always external to $r_\gamma$ \cite{Patil2012}.

As already anticipated, we proceed by working in a perturbative scheme. The coefficients $\alpha$ and $\beta$, as well as the expansion of $r_c$ up to order $(\gamma - 1)$, result to be
\begin{widetext}
    \begin{align}
        \alpha(r_0) ={}& \frac{r_0^{-2\gamma} \left[r_0^{\gamma} - (r_0 - 1)^{\gamma}\right] \left[E^2 r_0^{\gamma} (2r_0 - 2\gamma  - 1) + (\gamma - 2r_0 + 1) (r_0 - 1)^{\gamma}\right]}{\gamma E^2}, \label{alphaJNW} \\
        \beta(r_0) ={}& \frac{\left[\left(\frac{r_0 - 1}{r_0}\right)^{\gamma } - 1\right]^2 \left[\left(\frac{r_0 - 1}{r_0}\right)^{\gamma} - E^2\right] r_0^{\gamma} (r_0 - 1)^{-\gamma} \left\{3\gamma + 1 + 2\left[\gamma^2 - 3\gamma r_0 + r_0 (r_0 - 1)\right]\right\}}{2\gamma^2 E^2}, \label{betaJNW} \\
        r_{c}(x,\gamma) ={}& 1 + \frac{2}{3x + 1} + \frac{(\gamma - 1) (1 + x) \left[9x^2 + 6x \left(3 + \log\frac{9}{4} \right) + 12 (x - 1)\log(1+x) + 5 - 12 \log\frac{3}{2}\right]}{4x (1 + 3x)^2}. \label{rcJNW}
    \end{align}
\end{widetext}
The above relations reduce to Eqs.\ \eqref{alpha SCHW}, \eqref{beta SCHW} and \eqref{rcE} in the limit $q \to 0$, and to Eqs.\ (54) and (55) in Ref.\ \cite{Bozza2002} in the limit $E \to \infty$, as expected. From Eqs.\ \eqref{alphaJNW}, \eqref{betaJNW} and \eqref{rcJNW}, the strong deflection limit coefficients can be easily calculated, proceeding in exactly the same way as we did in the previous two sections. The deflection angle is then obtained as
\begin{multline} \label{deflangleJNWr0}
    \hat{\alpha}(r_{0},E,\gamma) = -a(x,\gamma)\log\delta(r_{0},x,\gamma) \\
    + b_{D}(x,\gamma) + b_{R}(x,\gamma) - \pi,
\end{multline}
where the coefficients $a(x,\gamma), b_{D}(x,\gamma)$ and $b_{R}(x,\gamma)$ are given by Eqs.\ \eqref{aJNW}, \eqref{bDJNW} and \eqref{bRJNW}--\eqref{bR1JNW} in the Appendix, respectively. We remind the reader that the quantity $x=x(E)$ is defined in Eq.\ \eqref{definition x}.
\begin{figure}[t]
\includegraphics[width=8cm]{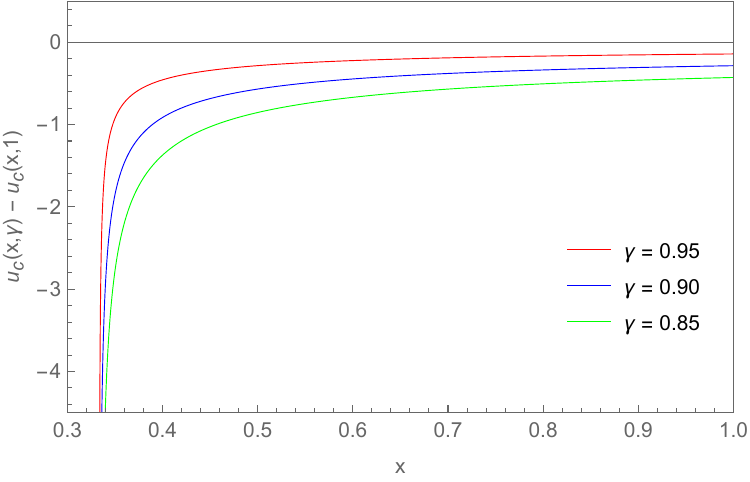}
\centering
\caption{Difference $u_c (x,\gamma) - u_c (x,1)$ as a function of $x$, for different values of the parameter $\gamma$. The quantity $u_c (x,\gamma)$ is defined in Eq.\ \eqref{ipcharged} (Sec.\ \ref{JNW}).}
\end{figure}
\begin{figure}[t]
\includegraphics[width=8cm]{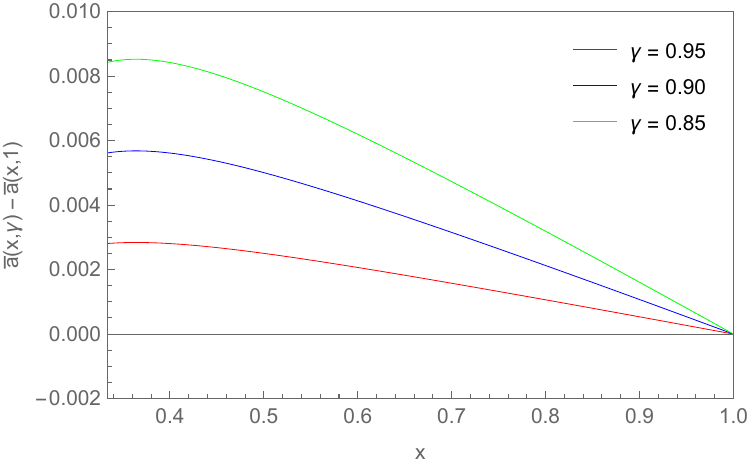}
\centering
\caption{Difference $\Bar{a}(x,\gamma) - \Bar{a}(x,1)$ as a function of the particle's energy, for different values of the parameter $\gamma$.\ The coefficient $\Bar{a}(x,\gamma)$ is defined by $a(x,\gamma)/2$, with $a(x,\gamma)$ given by Eq.\ \eqref{aJNW} (Sec.\ \ref{JNW}).}
\end{figure}
\begin{figure}[t]
\includegraphics[width=8cm]{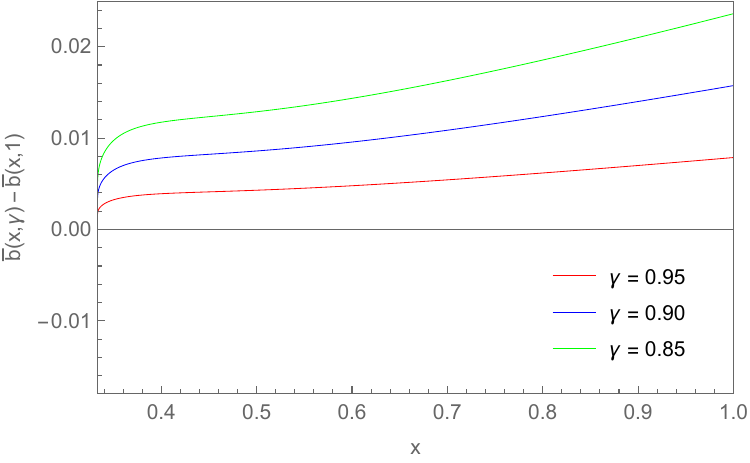}
\centering
\caption{Difference $\Bar{b}(x,\gamma) - \Bar{b}(x,1)$ as a function of $x$, for different values of the parameter $\gamma$. The coefficient $\Bar{b}(x,\gamma)$ is given by Eq.\ \eqref{bbarJNW} (Sec.\ \ref{JNW}).}
\end{figure}

In terms of the impact parameter, the deflection angle instead results in
\begin{equation} \label{deflangleJNW}
    \hat{\alpha}(u,E,\gamma) = -\Bar{a}(x,\gamma)\log\varepsilon(u,x,\gamma) + \Bar{b}(x,\gamma),
\end{equation}
where $\Bar{a}(x,\gamma) = a(x,\gamma)/2$, with $a(x,\gamma)$ given by Eq.\ \eqref{aJNW}, $\Bar{b}(x,\gamma)$ given by Eq.\ \eqref{bbarJNW}, and $\varepsilon(u,x,\gamma)$ expressed as
\begin{align}
    \begin{split}
    \varepsilon(u,x,\gamma) ={}& \frac{u}{u_{c}(x,\gamma)} - 1, 
    \end{split}\\
    \begin{split} \label{ipgamma}
    u_{c}(x,\gamma) ={}& \frac{3}{2}\sqrt{\frac{3}{9x^2 - 1}}\left[\frac{2\left(1 + x\right)^{\frac{3}{2}}}{\sqrt{1 + 3x}} + (3x + 5) \right. \\
    & \left. \times \left(\frac{1 + x}{1 + 3x}\right)^{\frac{3}{2}}\log \frac{3(1 + x)}{2}(\gamma - 1) \right].
    \end{split}
\end{align}
The differences $u_{c}(x,\gamma) - u_{c}(x,0)$, $\Bar{a}(x,\gamma) - \Bar{a}(x,0)$ and $\Bar{b}(x,\gamma) - \Bar{b}(x,0)$ as functions of $x$, fixed $\gamma$, are plotted in Figs.\ 9, 10 and 11, respectively. Comparing with Figs.\ 6, 7 and 8, we observe that the phenomenology arising from the Janis-Newman-Winicour spacetime differs fundamentally from that of the Reissner-Nordström spacetime, allowing for potential distinction. In Fig.\ 9, we observe that the coefficient $\Bar{a}(x,\gamma)$ approaches unity as $x \to 1$, in agreement with the findings in Ref.\ \cite{Bozza2002}. 

We note that, as $\gamma$ decreases, $u_{c}(x,\gamma)$ decreases, while the deflection coefficients $\Bar{a}$ and $\Bar{b}$ increase, similar to the Reissner-Nordström case. However, significant differences arise at different energies; in Fig.\ 7, we show that in Reissner-Nordström case, the increase in $\Bar{a}$ with charge affects all energies, whereas in Janis-Newman-Winicour case, this increase is much more pronounced at lower energies. This indicates that deflection of massive particles can help distinguish metrics that appear degenerate when studied using massless particles alone. We remind that the variable $x$, see Eq.\ \eqref{definition x}, is a monotonic function of $E$: small values of $E$ correspond to small values of $x$, and large values of $E$ correspond to large values of $x$.

\section{Discussion and conclusions}
\label{sec:conclusions}

This study extends the understanding of gravitational deflection of massive particles by very compact objects. By employing the strong deflection limit approximation, the research delves into the behavior of massive particles traveling along timelike geodesics, particularly when these particles approach a compact object at a radius close to that of the unstable circular orbit. This scenario, where particles circle the object before eventually escaping, mirrors the lensing effects observed for light, thereby extending the application of gravitational lensing principles to particles with mass.

The key contribution of this work is the development of a general solution for the deflection angle of massive particles in the strong deflection limit. Unlike previous studies that focused on specific spacetime metrics, this study provides a framework applicable to any static, asymptotically flat and spherically symmetric spacetime, making it relevant to a wider range of astrophysical scenarios.

In terms of the closest approach distance, the formula for the deflection angle is given by Eq.\ \eqref{deflection angle final}, while as a function of the impact parameter it is given by Eq.\ \eqref{deflection angle imppar}. Recall that, unlike light rays, the deflection of massive particles depends on two parameters rather than one. Here, we choose the particle’s energy at infinity per unit rest mass, $E$, as one of these parameters. Note that in most formulas, the energy appears only within the quantity $x = x(E)$, defined in Eq.\ \eqref{definition x}. Alternatively, one can use the velocity at infinity by applying substitution \eqref{eq:E-velocity}.

As an application, the formulas of Sec.\ \ref{sec:IV} have been specialized to the Schwarzschild, Reissner-Nordström and Janis-Newman-Winicour metrics.\ In terms of $r_{0}$, the deflection angle is given by Eq.\ \eqref{deflection angle mp} for Schwarzschild, Eq.\ \eqref{deflanglechargedr0} for Reissner-Nordström and Eq.\ \eqref{deflangleJNWr0} for Janis-Newman-Winicour. Alternatively, in terms of $u$, it is given by Eq.\ \eqref{daimp} for Schwarzschild, Eq.\ \eqref{deflanglecharged} for Reissner-Nordström and Eq.\ \eqref{deflangleJNW} for Janis-Newman-Winicour.

\section*{Acknowledgements} \label{sec:acknowledgements}

OYT acknowledges support from the ERC Advanced Grant ``JETSET: Launching, propagation and emission of relativistic jets from binary mergers and across mass scales'' (Grant No.\ 884631).\ OYT also thanks Valerio Bozza and his group for their hospitality during a visit to the University of Salerno.

\appendix

\onecolumngrid

\section{Strong deflection limit coefficients for massive particles in Janis-Newman-Winicour spacetime}

In this Appendix, we report the analytical expressions for the strong deflection limit coefficients in the case of the Janis-Newman-Winicour metric, see Sec.\ \ref{JNW}.

The coefficients $a(x,\gamma)$ and $b_{D}(x,\gamma)$ are obtained as
\begin{align}
        \begin{split} \label{aJNW}
        a(x,\gamma) ={}& \frac{\sqrt{1 + x} \left\{48x^2 + (\gamma - 1) (x - 1) \left[3x (3x - 4) + 12 \log\frac{3(x+1)}{2}  - 5\right]\right\}}{24\sqrt{2}x^{5/2}}, 
        \end{split}\\
        \begin{split} \label{bDJNW}
        b_{D}(x,\gamma) ={}& \sqrt{\frac{2(1 + x)}{x}}\left\{\log2 + (\gamma - 1)\left[\frac{(x - 1)\log2}{48x^2} \left(9x^2 - 12x - 5 + 12\log \frac{3(1 + x)}{2}\right) \right. \right.
        \\
        & \left. \left. \hspace{8.1cm} -\frac{3x - 3(1 + x) \log\frac{3(1 + x)}{2} +1}{3 x + 1}\right]\right\}.
        \end{split}
\end{align}
The coefficient $b_{R}(x,\gamma)$ is given by
\begin{equation} \label{bRJNW}
    b_{R}(x,\gamma) = -\hspace{0.5mm}2\sqrt{\frac{1 + x}{2x}} \log\left[\frac{\left(\sqrt{3x - 1} + \sqrt{6x}\right)^2}{24x}\right] + (\gamma - 1)\int_0^1 \mathrm{b}_{R,1}(z,x)dz,
\end{equation}
where, in turn, the integrand $\mathrm{b}_{R,1}(z,x)$ is expressed as
\begin{multline} \label{bR1JNW}
        \mathrm{b}_{R,1}(z,x) = \frac{\left(1 - x^2\right) \left[3x(3x - 4) + 12\log\frac{3(1 + x)}{2} -5\right]}{24\sqrt{2x^5 (1 + x)}z} 
        + \left\{z \left[3x \left(2 - z\right) - z\right]\left[12x (2 - z)(z - 1) + z(6 - 5z) \right. \right. \\
        \left. \left. \hspace{-0.85cm} + \hspace{1mm} 9x^2 z (z - 2)\right]\left[5 + 9x^2  + 6x \left(3 + \log\frac{9}{4}\right) - 12\log \frac{3}{2} + 12(x - 1)\log (1 + x) \right] \right. \\
        \left. \hspace{0.45cm} + \hspace{1mm} 3z^2 \left[3x (z - 2) + z\right]^2 \left\{x\left[9 x^2 - 3x\left(13 + \log 16\right) - 29 + 8 \log \frac{6561}{32}\right] + 64x\log (1 + x) \right. \right. \\
        \left. \left. \hspace{0.95cm} + \hspace{1mm} 12(x - 1)^2 \log 3(x+1) - 5 - 12\log 2\right\} + 24x \left[3x(2 - z) - z\right]\left\{\left[6x (z - 1)(5z^2 - 6z - 1) \right. \right. \right. \\
        \left. \left. \left. \hspace{1.35cm} + \hspace{1mm} 10 + 9x^2 z - 15z + 2z^2 (5z - 3)\right]\log \frac{3(1 + x)}{2} + (z - 1)(z + 3xz + 2)\left[z + 3x(1 + z) + 5\right]\right\}\right\} \\
        \times \left[(1 + x)(1 + 3x)\right]\left\{8 \sqrt{3}x \left[3x (z - 2) + z\right]^2 \left[z(1 + 3x\right]^2 \sqrt{z^2(1 + x)\left[3x (2 - z) - z\right]}\right\}^{-1}.
\end{multline}
Finally, the expression for the coefficient $\Bar{b}(x,\gamma)$ is
\begin{multline} \label{bbarJNW}
        \Bar{b}(x,\gamma) = -\pi + b_{R}(x,\gamma) + \frac{\sqrt{1 + x}}{2x}\left\{\log 6 + \frac{\gamma - 1}{1 + 3x}\left\{\left[24x \left(9x^2 + 11x + 3\right) - 12\right]\log \frac{3(1 + x)}{2}\right.\right. \\
        \left. \left. \hspace{2.3cm} - \hspace{1mm} 2\left\{5 + 9x\left[3 + x(6x^2 + 17x + 9)\right] \right\} + (x - 1)\log 6 \left\{(3x - 5)(1 + 3x)^2 \right. \right. \right. \\
        \left. \left. \left. + \hspace{1mm} 12 \left[(1 + 6x)\log3(1 + x) - 3x\log6(1 + x) - \log 2\right]\right\}\right\}\right\}.
\end{multline}

\twocolumngrid

\end{document}